\documentclass[twocolumn,nofootinbib,amsmath,prd,aps,superscriptaddress,
tightenlines,preprintnumbers,nobalancelastpage]{revtex4}

\usepackage{slashed}
\usepackage{color, verbatim}
\usepackage{latexsym}
\usepackage{amsmath}
\usepackage{graphicx}

\usepackage{hyperref}

\usepackage{amssymb}
\usepackage{graphicx}
\usepackage{bm}
\usepackage{hyperref}
\usepackage{adjustbox}

\usepackage{graphicx}
\usepackage{latexsym}
\usepackage{amsmath}
\usepackage{mathrsfs}
\usepackage{hyperref}
\usepackage{enumerate}
\usepackage{adjustbox}
\usepackage{amssymb}
\usepackage{slashed}

\def\bea{\begin{eqnarray}}
\def\eea{\end{eqnarray}}

\begin{document}

\newcount\hour \newcount\minute
\hour=\time \divide \hour by 60
\minute=\time
\count99=\hour \multiply \count99 by -60 \advance \minute by \count99
\newcommand{\mydate}{\ \today \ - \number\hour :00}

\preprint{IPPP/19/63}
\title{\Large Novel $B$-decay signatures of light scalars at high energy 
facilities}

\author{Andrew Blance$^a$$^b$, Mikael Chala$^a$$^c$, Maria Ramos$^d$ and 
Michael 
Spannowsky$^a$\\\vspace{0.4cm}
\it {$^a$Institute of Particle Physics Phenomenology, Physics Department, Durham 
University, Durham DH1 3LE, UK}\\[0.2cm]
\it {$^b$Institute for Data Science, Durham 
University, Durham DH1 3LE, UK}\\[0.2cm]
\it {$^c$CAFPE and Departamento de F\'isica Te\'orica y del Cosmos,
Universidad de Granada, E–18071 Granada, Spain}\\[0.2cm]
\it {$^d$Laborat\'orio de Instrumenta\c{c}\~ao e F\'isica Experimental de 
Part\'iculas, Departamento de F\'isica
da Universidade do Minho, Campus de Gualtar, 4710-057 Braga, Portugal}
}

\begin{abstract}
We study the phenomenology of light scalars of masses $m_1$ and $m_2$ coupling 
to heavy flavour-violating vector bosons of mass $m_V$. For 
$m_{1,2}\lesssim $ 
few GeV, this scenario triggers the rare $B$ meson decays $B_s^0\to 3\mu^+ 
3\mu^-$, $B^0\to 3\mu^+ 3\mu^-$, $B^+\to K^+ 3\mu^+ 3\mu^-$ and $B_s^0\to 
K^{0*} 3\mu^+ 3\mu^-$; the last two being the most important ones for 
$m_1\sim m_2$. None of these signals has been studied experimentally; 
therefore we propose analyses to test these channels at the LHCb. We 
demonstrate that the reach of this facility extends to branching ratios as small 
as $6.0\times 10^{-9}$, $1.6\times 10^{-9}$, $5.9\times 10^{-9}$ and $1.8\times 
10^{-8}$ for the aforementioned channels, 
respectively. For $m_{1,2}\gg \mathcal{O}(1)$ GeV, we show that slightly modified versions of 
current multilepton and multitau searches at the LHC can probe wide regions of 
the parameter space of this scenario. Altogether, the potential of the searches 
we propose outperform other constraints such as those from meson mixing.
\end{abstract}

\maketitle

\tableofcontents

\section{Introduction}

Searches for new physics in final states often considered as ``standard candles'', most notably in searches for supersymmetry (SUSY), have not 
provided any evidence of physics beyond the Standard Model (BSM) so far. This 
fact does not
necessarily disproves low energy SUSY or other popular BSM extensions~\cite{Chala:2018pbn}, such as 
composite Higgs models (CHM)~\cite{Kaplan:1983fs,Kaplan:1983sm}. However, it supports the 
search for new physics in radically new and still unexplored channels.

In this paper we focus on light singlet scalars $a_{1,2}$ that can be produced in rare decays of 
$B$ mesons
mediated by heavy flavour-violating vector bosons $V$. This scenario is especially  
motivated, as it arises naturally in non-minimal CHMs~\cite{Gripaios:2009pe,Vecchi:2013bja,Sanz:2015sua,Chala:2016ykx,Balkin:2017aep,DaRold:2019ccj}. ($V$ and 
$a_{1,2}$ can be seen as the counterparts of the $\rho$ and the pions in QCD.) 
Likewise, such vector boson can explain the apparent anomalies observed in tests 
of lepton flavour universality~\cite{Niehoff:2015bfa,Niehoff:2015iaa,Carmona:2015ena,Megias:2016bde,GarciaGarcia:2016nvr,Sannino:2017utc,Carmona:2017fsn,Chala:2018igk}. Moreover, the bounds on such vector boson are weakened when it decays into lighter composite resonances~\cite{Chala:2018igk}, such 
as the aforementioned scalars. Finally, also supersymmetric models can trigger 
similar decays, mediated by scalar and pseudoscalar sgoldstino particles~\cite{Demidov:2011rd}.

If, similarly to the Higgs boson, the scalars couple stronger to the muon than 
to the electron, processes such as $B_s^0\to a_1 a_2$ \textit{can} lead to four 
muon 
final states. To the 
best of our knowledge, the corresponding signal has been studied experimentally 
only at the LHCb~\cite{Aaij:2016kfs}; the most stringent limit being 
$\mathcal{B}(B_s^0\to 
2\mu^+ 2\mu^-) < 2.5\times 10^{-9}$.

However, there are different reasons to consider alternative $B$ meson decay 
modes. To start with, the partial width for $a_2 \to a_1 a_1$ can very easily 
dominate over the corresponding leptonic width. In this case, six muon final 
states rather than four muon ones are to be studied. 
And secondly, the 
scalars couple to the mediator as a vector current $\sim a_1\partial a_2$. When 
the latter is conserved, namely for $m_1\sim m_2$ (and in particular in the 
massless limit), the $B$ meson decay into such scalars vanishes. In other words, 
$\Gamma(B_s^0\to a_1 a_2)\sim (m_1^2-m_2^2)/m_B$. In this regime, one should 
rather explore three body decays of $B$ with emitted mesons. In this work we 
focus mostly on $B^+\to K^+ 3\mu^+ 3\mu^-$. (The inclusion of conjugate modes 
of charged decays is implied throughout the paper.)

We 
also extend previous works on this 
topic~\cite{Demidov:2011rd,Nelson:2013ula,Chala:2019vzu} 
by studying the regime of large scalar masses. In such regime, $a_{1,2}$ can no 
longer show up in rare decays of $B$ mesons. However, they can appear in 
decays of the vector mediator if it is at the TeV scale and therefore be produced 
in $pp$ collisions at the LHC. 

This article is organised as follows. In section~\ref{sec:framework}, we 
provide the Lagrangian relevant for our study and define the region of the 
parameter space of phenomenological interest.
In section~\ref{sec:lhcb} we focus on the regime $m_{1,2} \lesssim $ few 
Gev and provide analyses for the LHCb and estimate the reach for different $B$ 
decays. We do not circumscribe to any particular 
value of $m_{1,2}$, but rather scan over different values of these. In 
section~\ref{sec:lhc} we focus instead on the regime $m_{1,2} >$ few GeV and 
study the corresponding LHC signatures. 

Unless otherwise stated, all limits given in this article stand for 95\% CL.

We 
conclude in 
section~\ref{sec:conclusions}, while we dedicate Appendix~\ref{sec:app} to 
building a complete model that predicts definite values of several of the 
parameters that we scan over.

\section{Framework}\label{sec:framework}


Let us consider the Lagrangian of the SM extended with a heavy vector $V$, and two light 
scalars $a_1, a_2$. The relevant Lagrangian before electroweak symmetry 
breaking (EWSB) (in the basis in which up quark and lepton Yukawas are 
diagonal) reads
\begin{align}\label{eq:lag}\nonumber
 L =\, &\frac{1}{2}m_V^2 V_\mu V^\mu + \frac{1}{2}m_1^2 a_1^2 + \frac{1}{2}m_2^2 
a_2^2 + m_{12} a_2 a_1^2  + \cdots\\
&+ V^\mu \left[g_{12} a_1\overleftrightarrow{\partial_\mu} a_2 + 
g_{qq}(\overline{q_L}\gamma_\mu q_L + \text{h.c.})\right]~,
\end{align}
with $m_V \gg m_{1,2}$. The ellipsis stand for terms not relevant for this 
study. Without loss of generality, 
we assume $m_2 > m_1$. The scalars $a_{1,2}$ can be more naturally 
thought of as the real and imaginary components of a complex field $\Phi$; the 
Lagrangian being invariant under $\Phi\to \exp{(i\theta)}\Phi$ up to 
$\mathcal{O}(1-m_2/m1, m_{12})$. In the Appendix~\ref{sec:app} we match a 
concrete CHM to the Lagrangian above.

Assuming that $V$ interacts mostly with the third generation quarks, after EWSB 
it couples to $\overline{b_L}b_L$ and $\overline{t_L}t_L$ as well as 
$\overline{b_L}s_L+\text{h.c.}$ with strengths $\sim g_{qq}$ and 
\begin{equation}
g_{sb}\equiv g_{qq} V_{ts}^{CKM} V_{tb}^{CKM}\sim 0.04\, g_{qq}~,
\end{equation}
respectively. 

We distinguish two 
different regimes depending on the masses of the scalars: $1\,\text{GeV} 
\lesssim m_{1,2} \lesssim 4$ GeV (\textit{low mass regime}) and $m_{1,2} > 4$ 
GeV (\textit{high-mass regime}). Likewise, we consider two possible scenarios 
for the couplings of $a_{1,2}$ to the fermions. First, we assume that $a_{1,2}$ 
are muonphilic. As a second possibility,
we assume that they couple only to
the SM leptons and with Higgs-like strength, namely $\sim g_{1,2} y_\ell 
a_{1,2} 
\ell^+\ell^-$, with $y_\ell$ the SM Yukawa couplings and $g_{1,2}$ free 
dimensionless parameters and lepton independent. 

In the \textit{low-mass regime}, $a_1$ decays mostly into muons irrespectively 
of whether it is muonphilic or just leptophilic. In the \textit{high-mass 
regime}, it decays mostly into taus unless it is muonphilic.

Regarding the decay of $a_2$, if $m_2 > 2 m_1$, then $a_2$ can 
either decay into $a_1 a_1$ or into lepton pairs, depending on $m_{12}/g_2$:
\begin{align}
 \Gamma(a_2 \to \ell^+\ell^-) &= \frac{g_2^2 y_\ell^2}{8\pi} 
\left(1-\frac{4 m_\ell^2}{m_2^2}\right)^{3/2}m_2~,\\
 \Gamma(a_2 \to a_1 a_1) &= \frac{m_{12}^2}{8\pi m_2} 
\left(1-\frac{4m_1^2}{m_2^2}\right)^{1/2}~.
\end{align}
In what follows, we assume that $m_{12}/m_2\gg g_2 y_\ell$ in this regime, so 
that 
$\mathcal{B}(a_2\to a_1 a_1) \gg \mathcal{B}(a_2\to\ell^+\ell^-)$.
Note that this 
inequality holds almost trivially, since one expects $m_{12}\sim m_2$ whereas 
the Yukawas are tiny.

If instead $m_2 < 2 m_1$, $a_2$ can either 
decay into  pairs of leptons as before, or into $a_1 \ell^+ \ell^-$ with width
\begin{align}
 \Gamma(a_2\to &a_1 \ell^+ \ell^-) \sim \nonumber \\
 &\frac{(g_1 y_\ell)^2}{64 \pi^3 
 m_2^3} m_{12}^2 m_1^2 
\bigg(1+\frac{m_2}{m_1}\bigg)\bigg(\frac{m_2}{m_1}-1\bigg)^5~.
\end{align}
This decay mode dominates if 
$g_1\gtrsim 100 g_2$. We assume this hierarchy hereafter. Thus, for example for 
$g_1 =3$ and $g_2 = 
0.01$, $a_2$ decays always into four leptons mediated by $a_1$, 
which can be either on-shell or off-shell. Also, they both have widths smaller 
than $10$ MeV and lifetime shorter than $10$ fs. As a 
consequence, both $a_{1,2}$ would seem to have vanishing 
experimentally measurable widths and flight distances. Furthermore, note that the 
Yukawa suppression helps 
also avoiding bounds from BaBar and even the future 
Belle-II~\cite{Liu:2018xkx}.

\begin{figure}[t]
 \begin{center}
  \includegraphics[width=0.49\columnwidth]{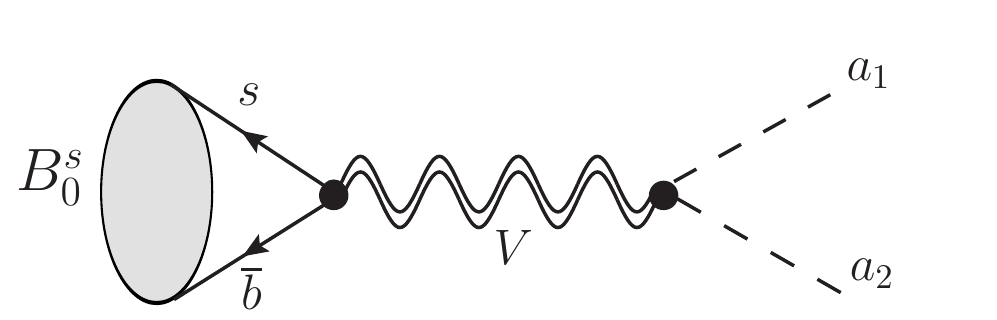}
  \raisebox{-0.3cm}{\includegraphics[width=0.49\columnwidth]{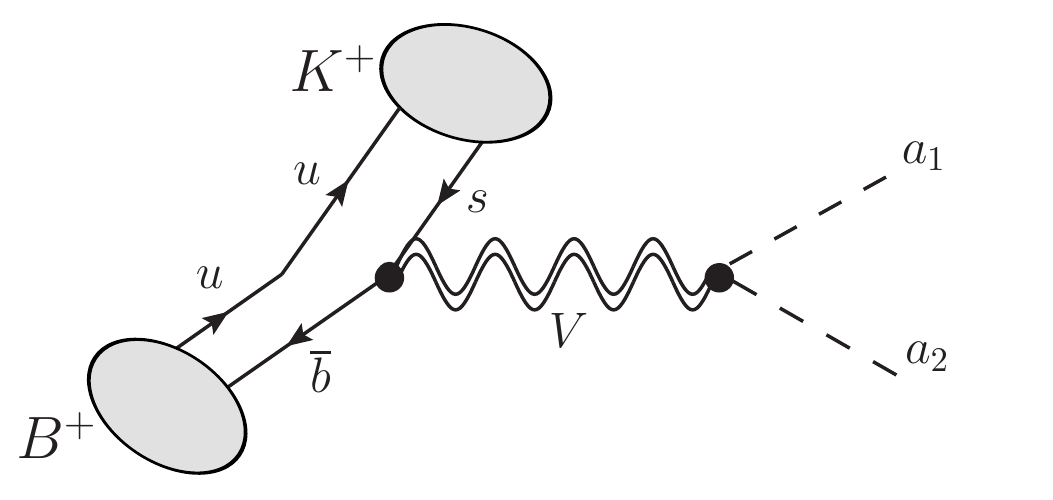}}
 \end{center}
 \caption{\it Tree level Feynman diagram for the 
decays $B_s^0\to a_1 
a_2$ (left) and $B^+\to K^+ a_1 a_2$ (right).}\label{fig:diagrams}
\end{figure}

At low energies, the vector boson $V$ triggers $B$ meson decays into the light 
scalars; see 
Fig.~\ref{fig:diagrams}. Depending on the relative size between $m_B$ and 
$m_{1,2}$ we distinguish two cases:

\begin{itemize}
 \item If $m_B > m_1+m_2$, we have $B_s^0\rightarrow a_1 a_2$~.
 \item If $m_B < m_1+m_2$ and $m_B > 3 m_1$, we have instead $B_s^0\to a_1 a_1 
a_1$. (Other three body decays, \textit{e.g.} $B_s^0\to a_1 \mu^+ \mu^-$ are 
subdominant due to the Yukawa suppression.)
\end{itemize}

\begin{figure}[t]
 \begin{center}
  \includegraphics[width=\columnwidth]{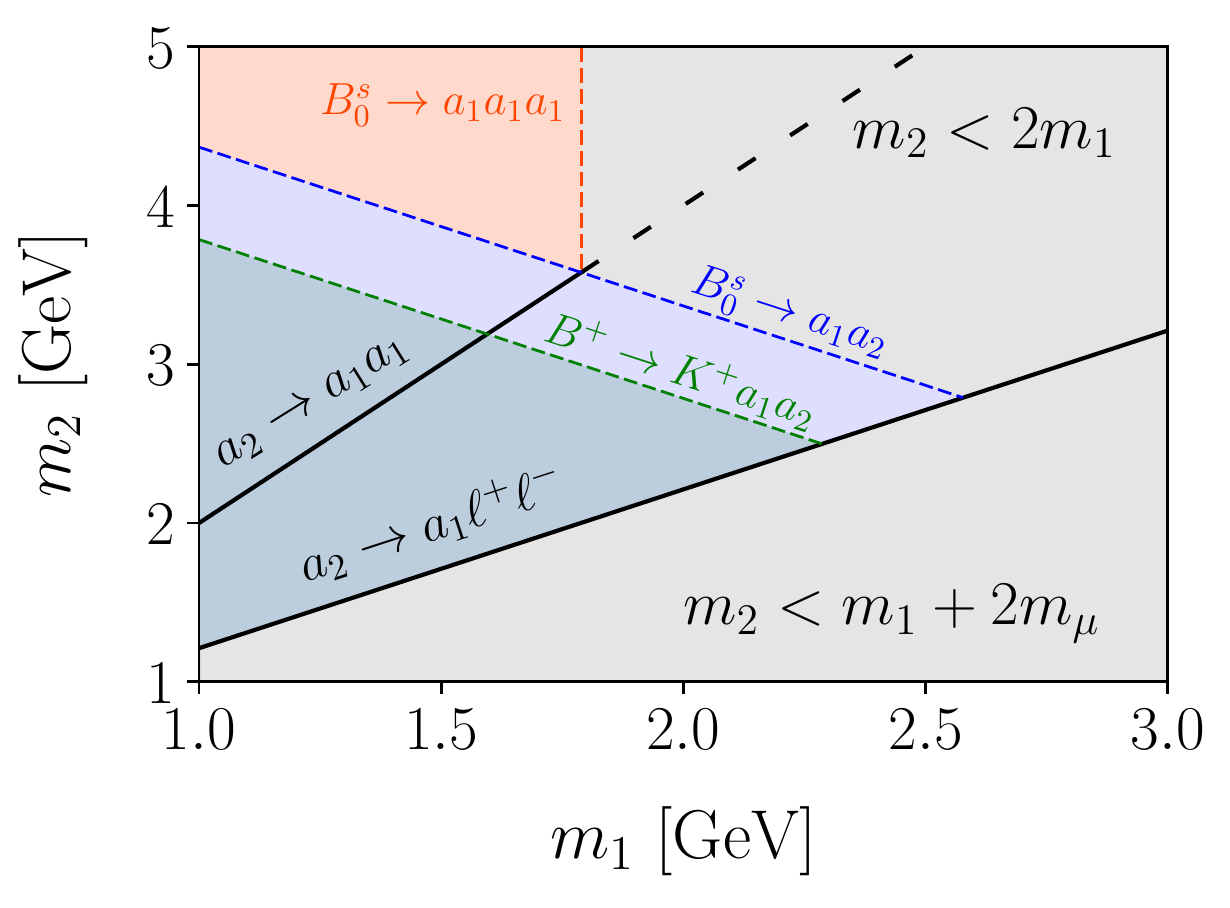}
 \end{center}
\caption{\it Dominant decays taking place in the different regions 
of the plane $(m_1, m_2)$. The gray areas are not 
considered in this analysis.}\label{fig:regions}
\label{fig:decays}
\end{figure}

If $m_B > m_1 + m_2 + m_K$, we also have $B^+\rightarrow K^+ a_1 a_2$. We do 
not consider any other cases in this paper; see Fig.~\ref{fig:regions}.

The decay width for $B_s^0 \rightarrow a_1 a_2$ reads:
\begin{equation}\label{eq:2bodywidth}
\Gamma = \frac{f_B^2}{16 \pi m_V^4} \left(g_{sb} g_{12}\right)^2
\frac{\left(m_1^2 - m_2^2\right)^2}{m_B}~\mathcal{K}\bigg(\frac{m_1}{m_B}, 
\frac{m_2}{m_B}\bigg)
\end{equation}
with
\begin{equation}
\mathcal{K}(x, y) = \bigg[x^4 + (1-y^2)^2 - 2x^2 (1+y^2)\bigg]^{1/2}
\end{equation}
and $f_B \sim 0.23$ GeV~\cite{Cheung:2006tm}.

The amplitude for $B_s^0\to a_1 a_1 a_1$ reads:
 \begin{equation}
 \mathcal{M} = 2 g_{sb} g_{12} \frac{f_B m_{12}}{m_V^2} \left[ \frac{q_{12}^2 - 
m_1^2}{q_{12}^2 - m_2^2} +  \frac{q_{23}^2 - 
m_1^2}{q_{23}^2 - m_2^2} +  \frac{q_{13}^2 - m_1^2}{q_{13}^2- m_2^2}\right]~,
 \end{equation}
where we have defined the transferred momenta $q_{12}^2 = (p_1 +  p_2)^2$,  $q_{23}^2 = (p_2 + p_3)^2$ and $q_{13}^2 = (p_1 + p_3)^2 = 3m_1^2 + m_B^2 - q_{12}^2 - q_{23}^2$. After integrating over $q_{23}^2$, we obtain: 
\begin{equation}
\frac{d \Gamma}{dq_{12}^2} = \frac{\left(g_{sb} g_{12}\right)^2 m_2^2}{384 \pi ^3 
m_B^3}\left(\frac{f_B 
m_{12}}{m_V^2}\right)^2 F\left[\frac{m_1}{m_2}, \frac{m_B}{m_2}, \frac{q_{12}}{m_2} , \frac{q_{23}}{m_2}\right]_{(q_{23}^2)^{\rm min}}^{(q_{23}^2)^{\rm max}}
\end{equation}
with
\begin{widetext}
\begin{align}
 F & (x, y,w,v) = \left(1-x^2\right)^2 \left[\frac{1}{1 - v^2} + \frac{1}{3x^2 + y^2 - w^2 - v^2 - 1} \right] + \frac{v^2 \left(2 + x^2 - 3 w^2 \right)^2}{(w^2 -1)^2} +  2(x^2 - 1) \times \nonumber \\
&\times \left\{ \frac{3 x^4 + x^2 (3+ y^2 - 9 w^2) + y^2 (2 - 3 w^2) + 3(w^4 + w^2 -1 )}{(w^2 - 1 ) (3x^2 + y^2 -w^2 -2)} \left[ \log{\left(v^2 - 1\right)} - \log{\left(1 + w^2 + v^2 - 3x^2 - y^2\right)} \right] \right\}
\end{align}
\end{widetext}
which should be evaluated at
\begin{equation}
\begin{aligned}
&\left(q_{23}^2\right)^{\rm max} = \left(E_2^*+ E_3^* \right)^2 - \left(\sqrt{E_2^{*2} - m_1^2} - \sqrt{E_3^{*2} - m_1^2} \right)^2 \\
&\left(q_{23}^2\right)^{\rm min} = \left(E_2^*+ E_3^* \right)^2 - \left(\sqrt{E_2^{*2} - m_1^2} + \sqrt{E_3^{*2} - m_1^2} \right)^2~,
\end{aligned}
\end{equation}
where $E_2^* \equiv q_{12}/2$ and $E_3^* \equiv \left(m_B^2 - q_{12}^2 - m_1^2\right)/(2q_{12})$. The final width is obtained integrating over $q_{12}^2$ between $4 m_1^2$ and $(m_B-m_1)^2$~.

In the limit $m_1, m_2 \rightarrow 0$, the integrated width simplifies to:
\begin{equation}
\Gamma \sim \frac{3\left(g_{sb} g_{12}\right)^2}{256 \pi^3} \frac{f_B^2 
m_{12}^2}{m_V^4}  m_B~.
\end{equation}

Finally, the amplitude for $B^+\to K^+ a_1 a_2$ is given by:
\begin{equation}
\label{eq: BK}
\mathcal{M} = -\frac {g_{s b} g_{1 2}}{m_V^2} \left< K(p_3) | \overline{s} \gamma_\mu b | B(p) \right> (p_2 - p_1)^\mu~,
\end{equation}
with
\begin{align}
\left< K(p_3) | \overline{s} \gamma_\mu b | B(p) \right> = & f_+ (q^2) \bigg[ (p + p_3)_\mu - \frac{m_B^2 - m_K^2}{q^2} q_\mu \bigg] \nonumber \\
& + f_0 (q^2) \frac{m_B^2 - m_K^2}{q^2} q_\mu
\end{align}
and again $q^2 = (p - p_3)^2$ is the transferred momentum, ranging from $(m_1 + 
m_2)^2 < q^2 < (m_B - m_K)^2$. The contraction 
of this matrix element with $(p_2 - p_1)$ in Eq.~\ref{eq: BK} simplifies to
\begin{align}
  \mathcal{M} &=   - \frac {g_{s b} g_{1 2}}{m_V^2} \bigg\lbrace \frac{\left(m_B^2 - m_K^2\right) \left(m_2^2 - m_1^2\right)}{q^2} \left[f_0 (q^2) - f_{+} (q^2)\right]  \nonumber \\ 
&  + \left[2\left(p_2 + p_3\right)^2  + q^2 - m_1^2 - m_2^2 - m_B^2 - m_K^2\right] f_{+} (q^2)\bigg\rbrace~.
\end{align}
For convenience, we trade these variables for $M_{12}^2 \equiv m_2^2 - m_1^2$ and $M_{BK}^2 
\equiv m_B^2 - m_K^2$, getting
\begin{equation}
\frac{d \Gamma}{d q^2} = \frac{\left( g_{sb} g_{12}\right)^2}{768 \pi^3 m_V^4 m_B^3} F(q^2)~,
\end{equation}
with
\begin{widetext}
\begin{align}
F (q^2)   & =   \frac{1}{q^2}\left[\frac{(M_{BK}^2 + q^2)^2}{q^4} - 4 
\frac{m_B^2}{q^2}\right]^{1/2} \left[\frac{(M_{12}^2 + q^2)^2}{q^4} - 
4\frac{m_2^2}{q^2}\right]^{1/2}	\nonumber \\
 &\times\bigg\{3  M_{BK}^4 M_{12}^4 |f_0 (q^2)|^2 +  \left[ q^4+ 2q^2 \left(M_{BK}^2 - 2 
m_B^2\right) + M_{BK}^4 \right] \left[ q^4 + 2 q^2 \left( M_{12}^2 - 2m_2^2\right) + 
M_{12}^4\right] |f_{+} (q^2)|^2 \bigg\}~.
\end{align}
\end{widetext}
Following Ref.~\cite{Ball:2004ye}, 
we parameterize the form factor as
\begin{equation}
f_+ (q^2) = \frac{r_1}{(1- q^2/m^2)} + \frac{r_2}{(1-q^2/m^2)^2}~,
\end{equation}
with $r_1 = 0.162$, $r_2 = 0.173$ and 
$m^2 = 5.41^2$ $\rm{GeV}^2$. Similarly,
\begin{equation}
f_0 (q^2) = \frac{r_2}{(1-q^2 /m_{\rm fit}^2)} 
\end{equation}
with $r_2 = 0.330$ and $m_{\rm fit}^2 = 37.46$ $\rm{GeV}^2$.
Finally, in the approximation $m_1, m_2, m_K \rightarrow 0$ and $f_0 , f_+ (q^2) \rightarrow 1$, we obtain:
\begin{equation}
\Gamma \sim  \frac{\left(g_{sb} g_{12}\right)^2}{3072 \pi^3 m_V^4}  m_B^5~.
\end{equation}

\begin{figure}[t]
 \begin{center}
  \includegraphics[width=\columnwidth]{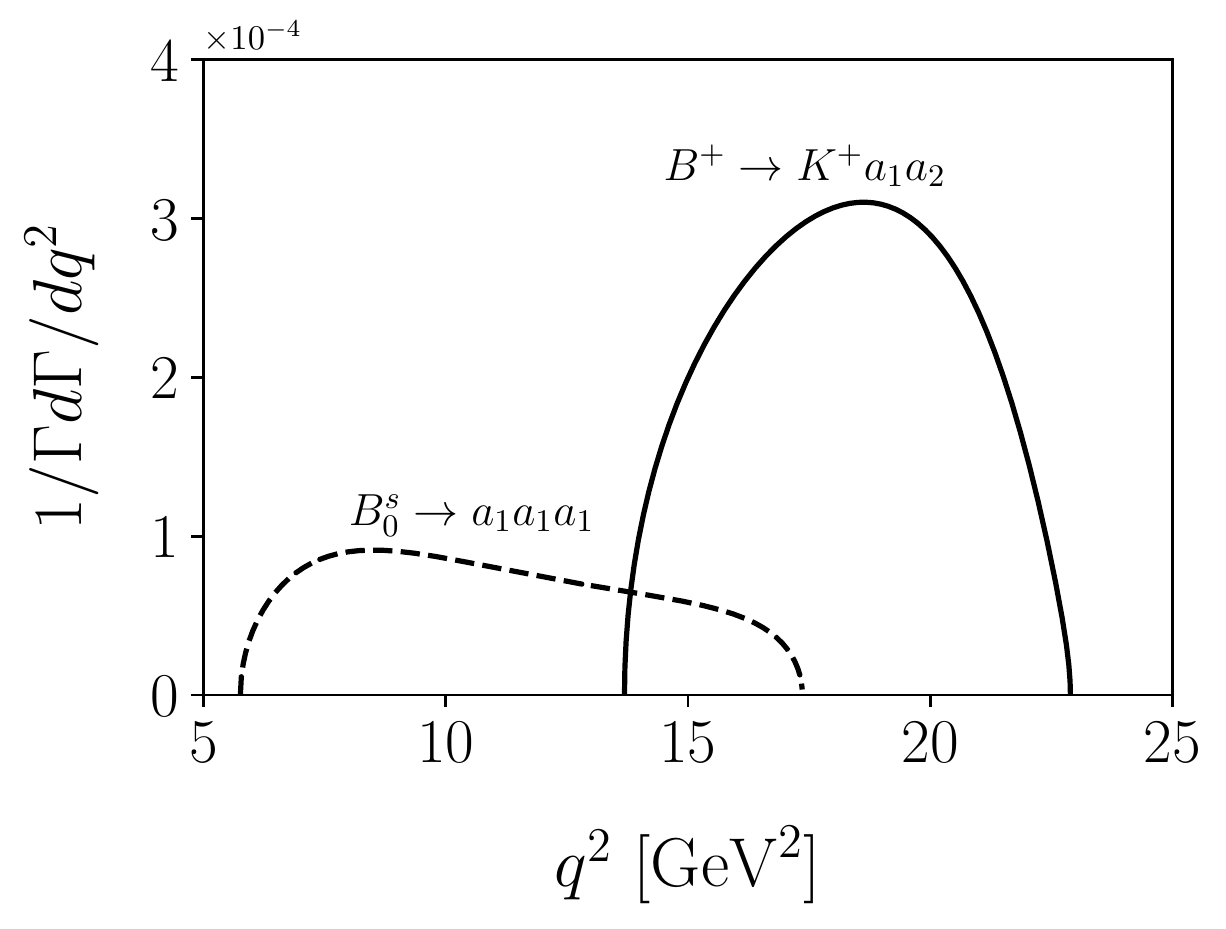}
 \end{center}
\caption{\it Differential branching ratios as a function of  $q^2=(p_1+p_2)^2$. We have 
fixed 
$m_1 = 1.2$ GeV, $g_{sb} = g_{12} = 1$, $m_V = 1$ TeV and $m_{12} = 5$ GeV. 
Due to the different kinematic regions where these decays take place, we have 
set $m_2 = 2.5$ GeV and $m_2 = 5$ GeV for the $B^+ \rightarrow K^+ a_1 a_2$ and $B_0^s \rightarrow a_1 a_1 a_1$, 
respectively. }
\label{fig:diff_widths}
\end{figure}

In Fig.~\ref{fig:diff_widths}, we show the magnitude of three body decays under 
consideration and their dependence with $(p_1+p_2)^2$. 
\begin{figure}[t]
 \begin{center}
  \includegraphics[width=\columnwidth]{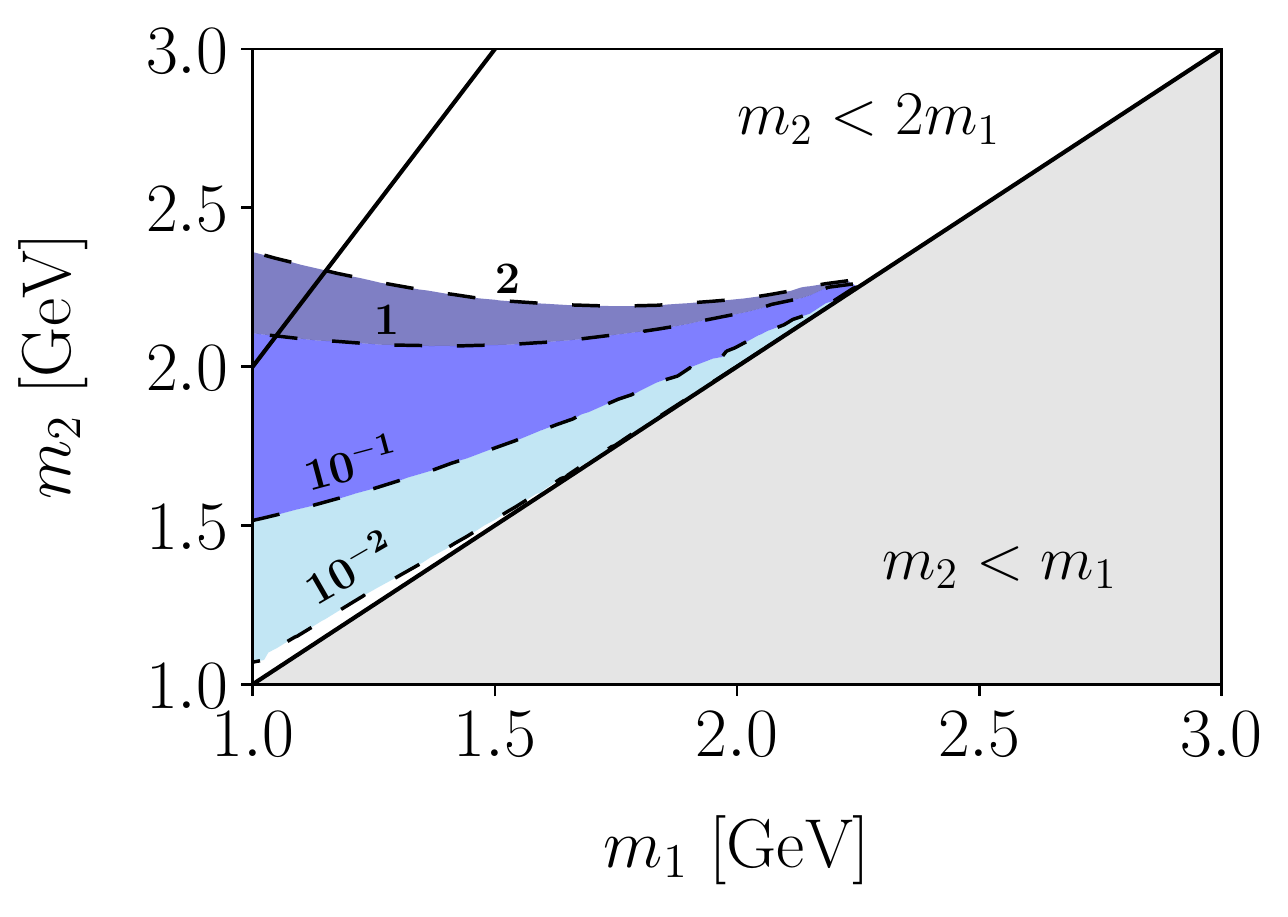}
 \end{center}
\caption{\it Value of 
$\Gamma \left(B_s^0 \rightarrow a_1 a_2 \right)/\Gamma \left(B^+ \rightarrow K^+ 
a_1 a_2 \right)$ in the plane $(m_1, m_2)$. This ratio vanishes along the line $m_1 = m_2$. }
\label{fig:ratios}
\end{figure}
In Fig.~\ref{fig:ratios}, we show the ratio of $\Gamma(B_s^0\to a_1 a_2)$ to 
$\Gamma(B^+\to K^+ a_1 a_2)$. It is very worth noting that it vanishes in the 
limit $m_1\to m_2$; see also Eq.~\ref{eq:2bodywidth}. In this regime, searches 
for $B_s^0$ decaying only to muons are irrelevant; extra mesons have to be 
tagged instead. There are however no analyses (not even prospects) in this 
respect, and this is a gap that we try to overcome in this work.

At high energies, $V$ can be produced on-shell in $pp$ collisions initiated by 
bottom quarks, and subsequently decay into third generation quarks and into 
$a_1 a_2$ with respective widths:

\begin{align}\nonumber
 \Gamma(V\to q\overline{q}) = \frac{g_{qq}^2}{8\pi} 
\left(1-\frac{m_q^2}{m_V^2}\right)\left(1-\frac{4m_q^2}{m_V^2}\right)^{1/2} 
m_V~,
\end{align}
\begin{align}
\Gamma(V\to a_1 a_2) = 
\frac{g_{12}^2}{48\pi}\bigg[
1&-2\frac{m_1^2+m_2^2}{m_V^2}\\\nonumber
&+\frac{(m_2^2-m_1^2)^2}{m_V^4}\bigg]^{3/2}
m_V~,
\end{align}
with $q = t,b$. Note that the scalar decay mode dominates already for 
$g_{12}\gtrsim 3 g_{qq}$.


\section{Low mass regime at the LHC$\mathbf b$}\label{sec:lhcb}

\begin{table}[t]
\begin{center}
\begin{tabular}{|l|c|c|c|}
    \hline
    & \multicolumn{2}{|c|}{$m_X \geq m_1 + m_2$} & $m_X < m_1 + m_2$\\
    \hline
     & $m_2 \geq 2m_1$ & $m_2 < 2m_1$ & $m_X\geq 3m_1$\\
\hline
    \hline
    $B_s^0\rightarrow 3\mu^+ 3\mu^-$ & [0.02,0.03] & 
[0.01,0.02]  & [0.02,0.03]\\\hline
\hspace{0.6cm}limit ($\times 10^{-9}$)& $~[6.7,11.6]$ & 
$[7.9,18.2]$  & $[6.0,11.9]$\\\hline

    \hline
    $B^+\rightarrow K^+ 3\mu^+ 3\mu^- $  & [0.007,0.009]
& [0.003,0.009] & four-body\\\hline
\hspace{0.6cm}limit ($\times 10^{-9}$)& $~ [5.9,8.0]$ & 
$[6.0,16.6]$  & four-body\\\hline
\end{tabular}
\end{center}
\caption{\it Maximum and minimum efficiencies for selecting 
signal events in the channels $B_s^0\to 3\mu^+ 3\mu^-$ ($m_X = m_{B_s^0}$) and 
$B^+\to K^+ 3\mu^+ 3\mu^-$ ($m_X = 
m_{B^+} - m_{K^+}$) in each kinematic region. The 
upper limits ($\times 10^{-9}$) on the corresponding branching ratios for $3$ fb$^{-1}$ 
of data are also shown. We vary
$m_{1,2}$ in the coloured region of Fig.~\ref{fig:decays}, with $m_2 < 10$ GeV 
and $m_1 \geq 1.1$ GeV. (For smaller values of $m_1$ the efficiency is negligible.)
}
\label{tab:BRs}
\end{table}


In the low mass regime, the smoking gun signature of the Lagrangian in 
Eq.~\ref{eq:lag} is rare decays of $B$ mesons into final states containing six 
muons (and possibly other lighter mesons). Let us focus first on the channel 
$B_s^0\to 3\mu^+ 3\mu^-$. As we have already commented, there are no searches 
for this decay mode, and so neither constraints nor any direct way to 
estimate the potential of the LHCb to test this process. We therefore 
suggest the first analysis in this respect.

We first require events with at least one muon with 
$p_T > 1.7$ GeV; this cut ensures that the events pass the same hardware trigger 
used at $\sqrt{s} = 8$ TeV~\cite{Aaij:2016kfs}. We subsequently require exactly 
six muons, 
with vanishing total charge. We also require all muon tracks to have $p_T > 
0.5$ GeV and $2.5< \eta < 5.0$. Finally, we require all muons tracks to have 
total momentum larger than $2.5$ GeV to simulate the threshold for muon 
identification based on the penetration power through absorption plates in the 
detector.

Due to the six muons in the final state, the SM backgrounds are negligible to 
very good approximation. They arise mostly from resonant production of $J/\Psi$ 
and $\varphi$ with subsequent decays into muons; we completely remove them by 
enforcing that no zero charge muon pair has an invariant mass in the range 
$[0.95,1.09]\cup[3.0,3.2]$ GeV. (We 
lose sensitivity to signal events with $m_1$ in that region, though.)  Even searches for four muons are background free~\cite{Aaij:2016kfs,Chala:2019vzu}, so it is guaranteed 
that any observed event in the six lepton final state is due to the signal.

\begin{figure}[t]
 \begin{center}
  \includegraphics[width=\columnwidth]{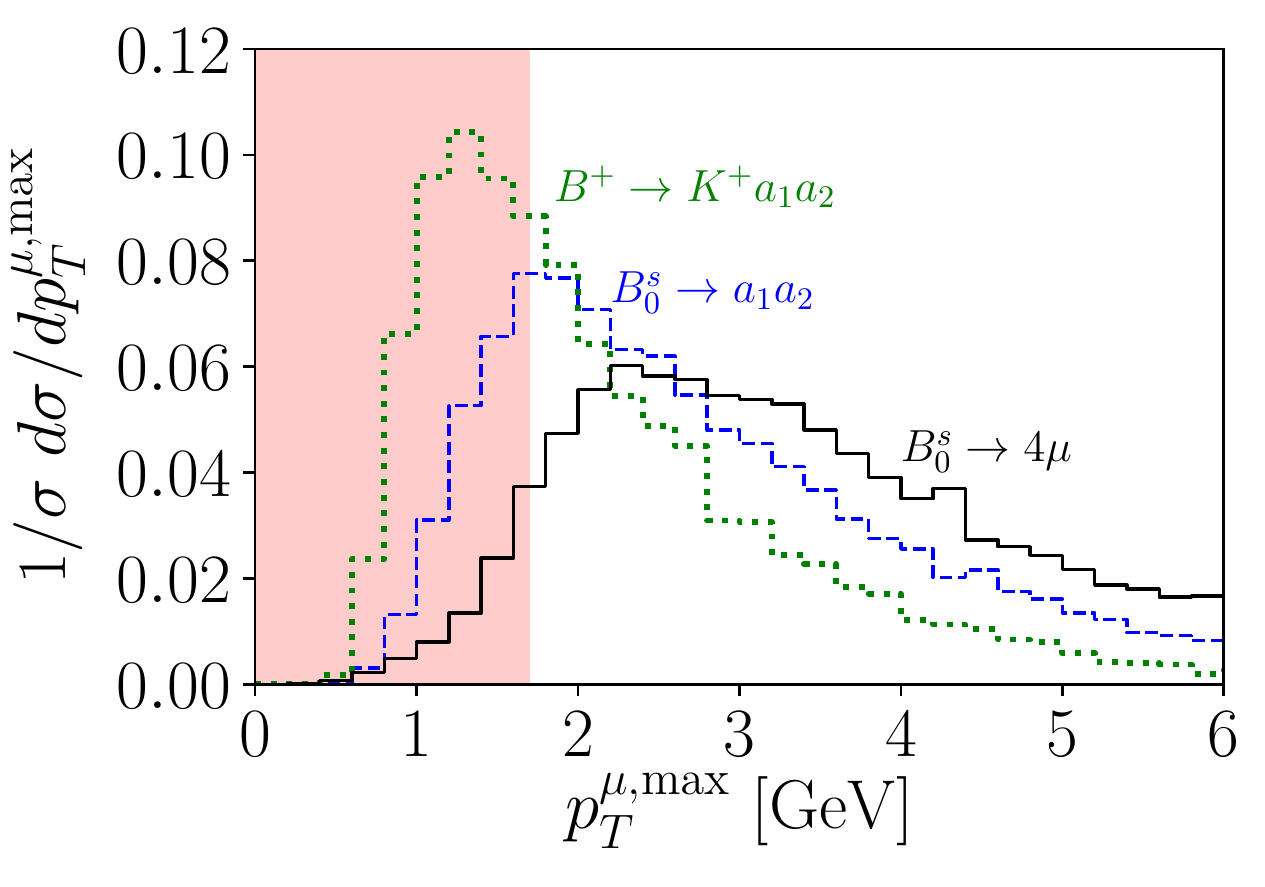}
 \end{center}
\caption{\it Normalized distribution of the transverse momentum 
of the hardest muon for $B_s^0 \rightarrow a_1 a_2$ and $B^+ 
\rightarrow K^+ a_1 a_2$ with $m_1 = 1$ GeV and $m_2 = 2.5$ GeV.  These 
distributions are compared with
the case $B_0^s \rightarrow 2\mu^+ 2\mu^-$.
}
\label{fig:max_pt}
\end{figure}

\begin{figure}[t]
 \begin{center}
  \includegraphics[width=\columnwidth]{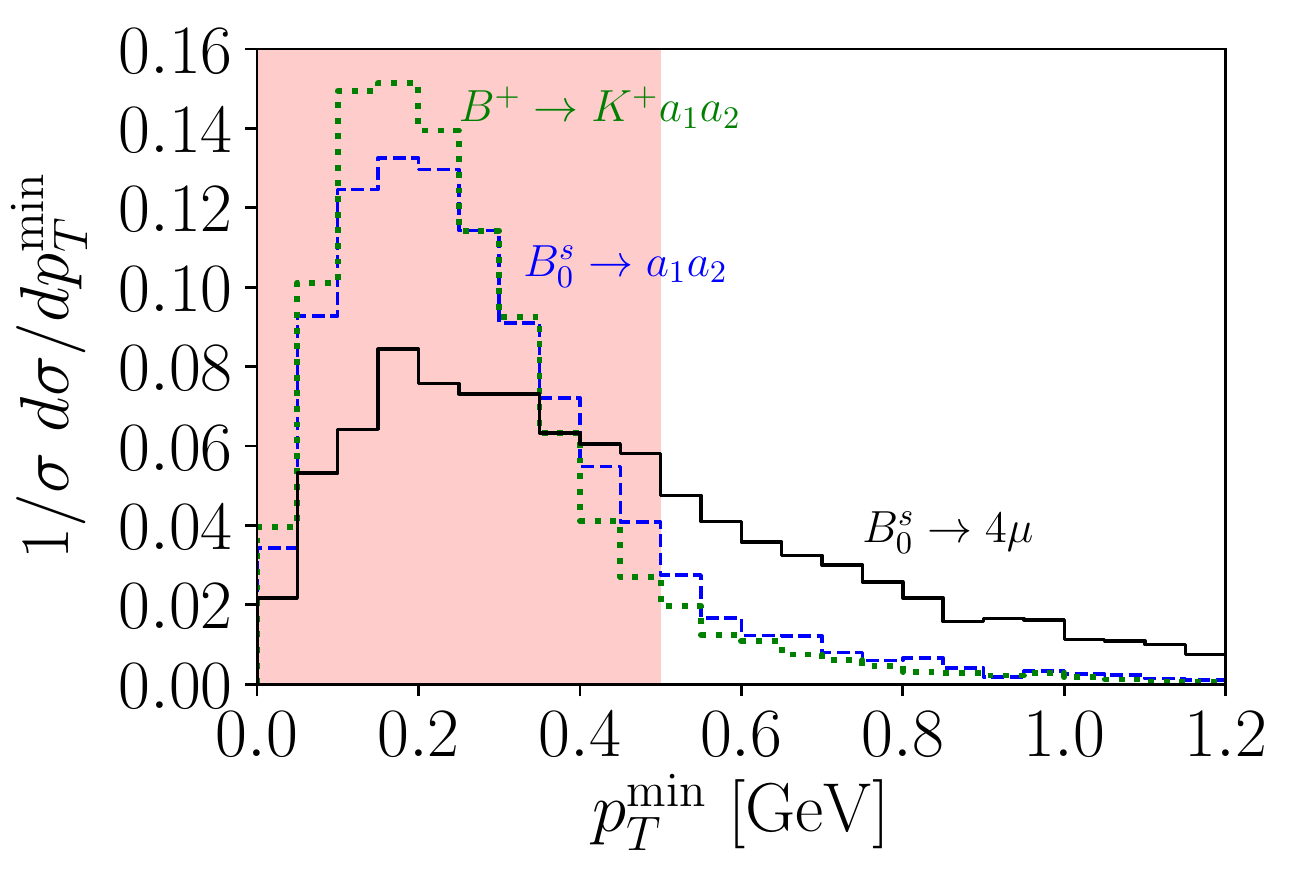}
 \end{center}
\caption{\it Normalized distribution of the transverse momentum of 
the softest track for $B_s^0 \rightarrow a_1 a_2$ and $B^+ 
\rightarrow K^+ a_1 a_2$ with $m_1 = 1$ GeV and $m_2 = 2.5$ 
GeV. These 
distributions are compared with
the case $B_0^s \rightarrow 2\mu^+ 2\mu^-$.
}
\label{fig:min_pt}
\end{figure}

We generate signal $B$ meson events using \texttt{Pythia 
v8}~\cite{Sjostrand:2014zea}; and 
\texttt{MadGraph v5}~\cite{Alwall:2014hca} with \texttt{Feynrules 
v2}~\cite{Degrande:2011ua} 
for the decays. (We have 
cross checked our event distributions using 
\texttt{EvtGen}~\cite{Lange:2001uf}.)
Following Ref.~\cite{Chala:2019vzu}, we compare the (mass dependent) 
efficiencies for 
selecting events in the channel $B_s^0\to 3\mu^+ 3\mu^-$ with that for 
$B_s^0\rightarrow 2\mu^+ 2\mu^-$. The former is 
shown in Tab.~\ref{tab:BRs}, while we estimate the latter to be 
$\varepsilon_{2\mu^+ 2\mu^-}\sim 0.14$.
The explanation for the smaller 
efficiencies for the six muon process is two fold. First, due to the larger 
number of final state tracks, there are more events with no single muon 
with $p_T > 1.7$ GeV which therefore do not pass the trigger; see 
Fig.~\ref{fig:max_pt}. And second, there are more muons with at least one track 
with $p_T < 0.5$ GeV which is therefore not detected; see 
Fig.~\ref{fig:min_pt}.

Given the absence of 
background, we can estimate the upper 
limit on the branching ratio of the new processes at $\sqrt{s} = 14$ TeV and  
luminosity 
$\mathcal{L}'$ as 
\begin{equation}
\mathcal{B}_{\text{max}}^{3\mu^+3\mu^-} \sim 
\frac{\mathcal{B}_{\text{max}}^{2\mu^+ 2\mu^-} \times 
\varepsilon_{2\mu^+2\mu^-} }{1.8 \times 
\varepsilon_{3\mu^+ 3\mu^-}} \times \frac{\mathcal{L}}{\mathcal{L}'}~,
\end{equation}
where $\mathcal{B}_{\text{max}}^{2\mu^+ 2\mu^-}$ is the upper limit on 
$\mathcal{B}\left(B_s^0 \rightarrow 
2\mu^+ 2\mu^-\right) = 2.5 \times 10^{-9}$, obtained in Ref.~\cite{Aaij:2016kfs} 
with 
$\mathcal{L} = 3~\rm{fb}^{-1}$ and 
$\sqrt{s} = 8$ TeV, under the same trigger and reconstruction criteria. The 
factor $1.8$ stands for the approximated growth of the $b$ production cross 
section from $\sqrt{s}=8$ TeV to $\sqrt{s} = 14$ TeV. The prospective bounds on 
the branching ratio of this new decay mode are given in Tab.~\ref{tab:BRs}.

We also consider the channel $B^+\to K^+ 3\mu^+ 3\mu^-$. In this case, on top 
of the selection criteria proposed before, we require the presence of a 
charged kaon which is also required to have $p_T > 0.5$ GeV and 
$2.5<\eta < 5.0$. The corresponding efficiencies are shown in 
Tab.~\ref{tab:BRs}. The limit on the branching ratio can be again obtained as

\begin{equation}
\mathcal{B}_{max}^{3\mu^+3\mu^- K^+} \sim \frac{\mathcal{B}_{max} 
^{2\mu^+ 2\mu^-} \times \varepsilon_{2\mu^+2\mu^-}}{1.8 
\times 3.7  \times \varepsilon_{3\mu^+ 3\mu^- K^+}}\times 
\frac{\mathcal{L}}{\mathcal{L}'}~,
\end{equation}
where the factor $3.7$ stands for the larger $B^+$ production cross 
section~\cite{Aaij:2011jp}. The bounds obtained 
this way are also shown in Tab.~\ref{tab:BRs}. It is worth noting that the 
prospective limits on this channel are comparable or even more stringent than that on the decay mode 
without the extra meson (due mostly to the larger cross section, that 
compensates the smaller efficiency). This fact, together with the observation 
that theoretically this decay mode dominates for $m_2\sim m_1$, strongly 
motivates searches for $B^+\to K^+ 3\mu^+ 3\mu^-$.

\begin{figure}[t]
 \begin{center}
  \includegraphics[width=\columnwidth]{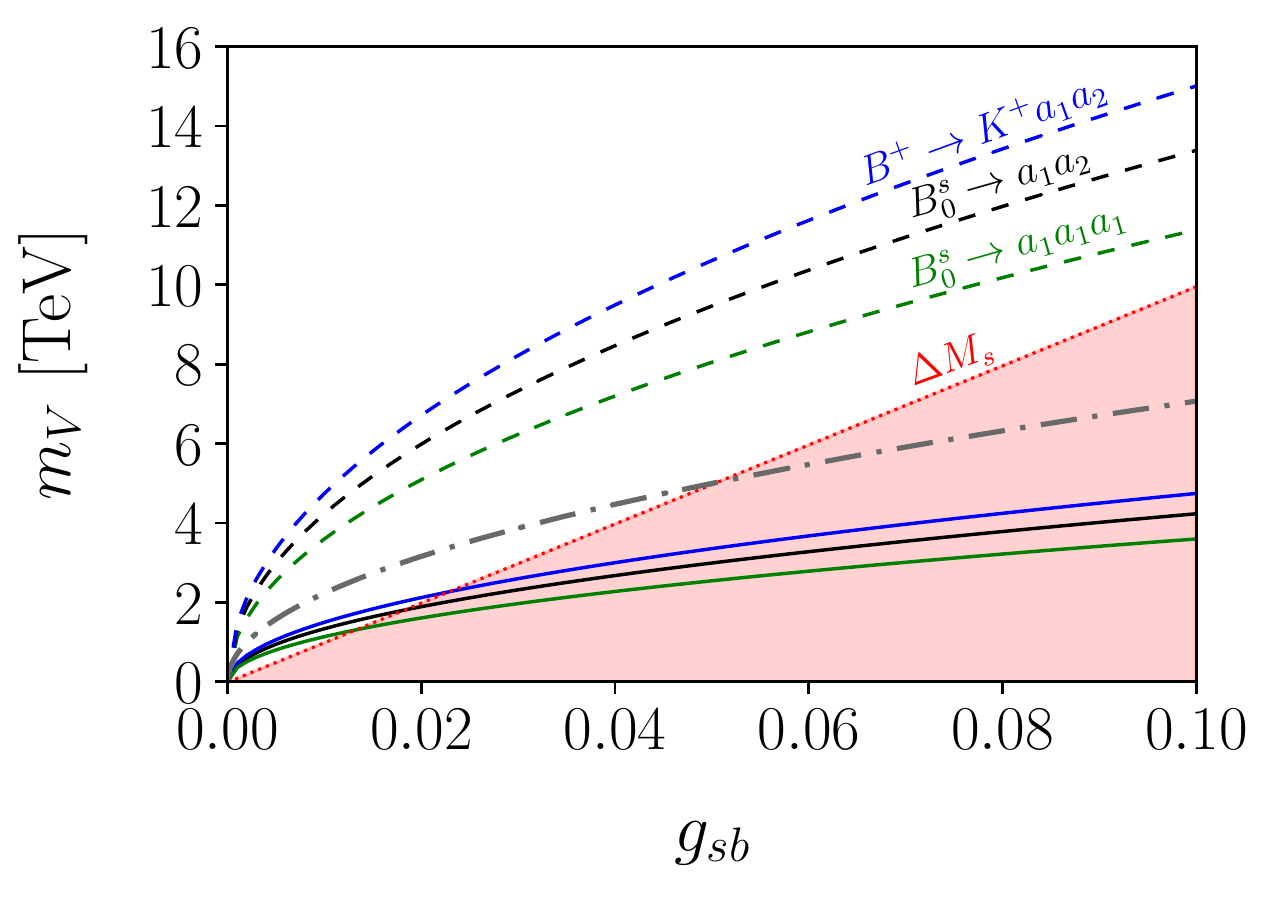}
 \end{center}
\caption{\it Maximum value of $m_V$ that can be tested in the 
searches for $B_s^0\to3\mu^+ 3\mu^-$ and $B^+\to K^+ 3\mu^+ 3\mu^-$ 
at the current run of the
LHCb (solid lines) and for Upgrade II (dashed lines). The red dotted line 
delimits the area excluded by measurements of $\Delta M_s$. In the  
dash-dotted line the anomalies in $R_K$ and $R_{K^*}$ can be explained at the $1\sigma$ level 
assuming $g_{V \ell \ell } \sim 1$~\cite{DiLuzio:2017fdq}. We have fixed $g_{12} = 0.5$ as 
well as $m_1 = 1.2$ GeV. We have set $m_2 = 2.0$ GeV for both $B_0^s 
\rightarrow a_1 a_2$ and $B^+ \rightarrow K^+ a_1 a_2$. For $B_0^s 
\rightarrow a_1 a_1 a_1$, we have fixed instead $m_2 = m_{12} = 5$ GeV.}
\label{fig:curr_bound}
\end{figure}

For illustration, we translate the expected limits in Tab.~\ref{tab:BRs} to the 
plane $(g_{sb}, m_V)$ in Fig.~\ref{fig:curr_bound} for definite values of 
$g_{12}$, $m_1$, $m_2$ and $m_{12}$ (when relevant). Prospects for 
the Upgrade II, defined by $\mathcal{L}' = 300$ fb$^{-1}$, are also shown. 
It is interesting to see that with our proposed analyses we can 
easily test masses larger than 15 TeV, thereby outperforming constraints obtained 
from $\Delta M_s$ and completely probing the region in which the anomalies in 
lepton flavour universality can be explained.

\begin{figure}[ht!]
 \begin{center}
  \includegraphics[width=\columnwidth]{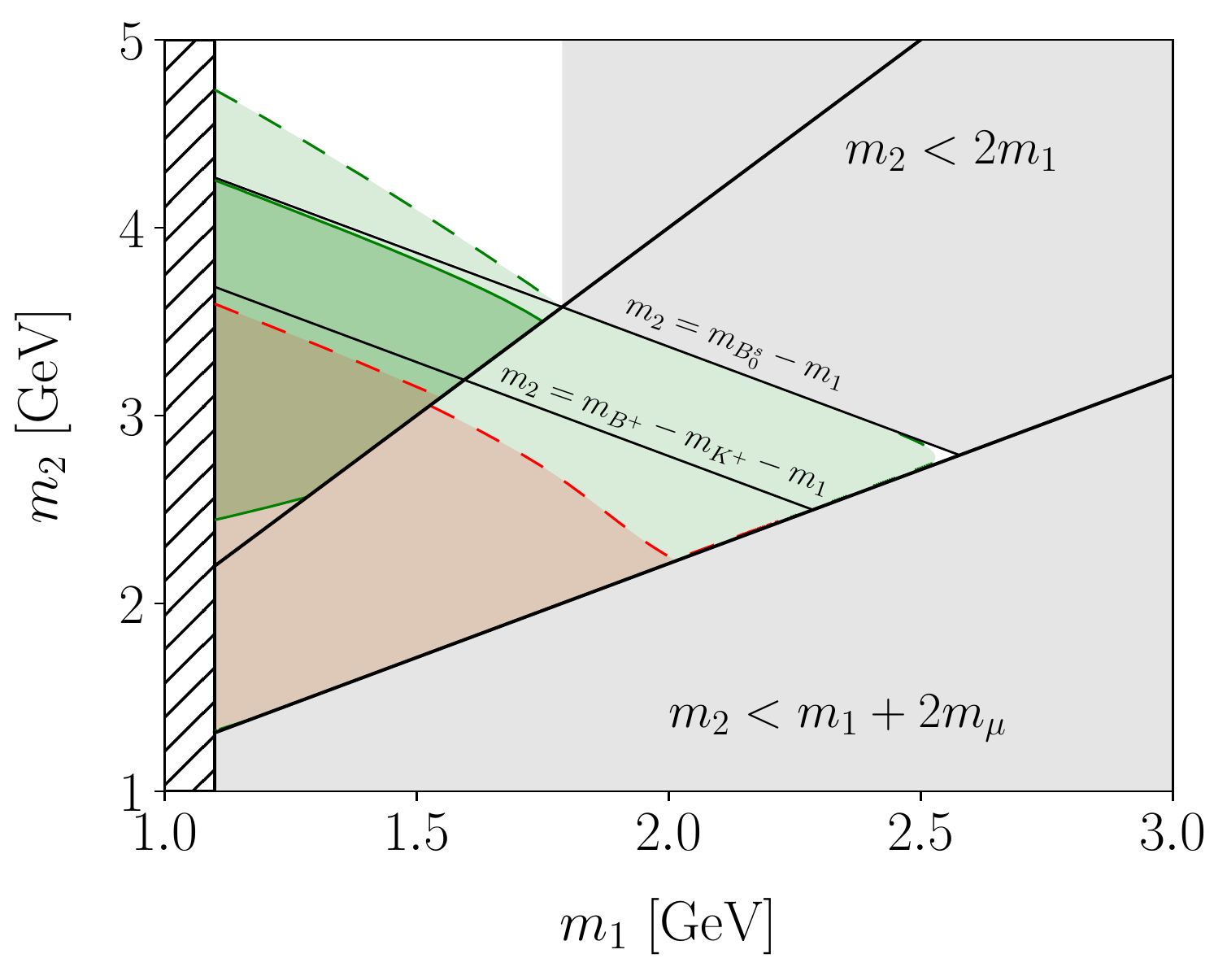}
 \end{center}
\caption{\it Region of the plane $(m_1, 
m_2)$ that can be tested at the current 
run of the LHCb (solid) and in Upgrade II (dashed) in searches for $B_s^0\to 
3\mu^+ 3\mu^-$ (green) and $B^+\to K^+ 3\mu^+ 3\mu^-$ (red). We 
have fixed 
$g_{sb} = 0.04$, $m_V = 4$ TeV and $g_{12} = 0.5$, as well as $m_{12} = 1$ 
GeV (only relevant in the upper left region). The sensitivity is negligible in the slashed region.
}
\label{fig:upgII_bound}
\end{figure}

Likewise, we also translate the aforementioned bounds to the plane $(m_1, m_2)$ 
in Fig.~\ref{fig:upgII_bound},
fixing $g_{sb} = 0.04$ as well as
$m_V = 4$ TeV. Such values are not yet excluded by measurements of $\Delta 
M_s$; see Refs.~\cite{DiLuzio:2017fdq}. In both figures, only the weakest limits of Tab.~\ref{tab:BRs} are used.


We also note that, if a signal is observed in these six-muon channels, the mass 
of the scalar particles involved could be reconstructed due to the outstanding 
detector resolution of the LHCb.  To this aim, we provide two 
different algorithms, depending on whether $m_2 > 2 m_1$ (in which case $a_2\to 
a_1 a_1$) or rather $m_2 < 2m_1$ (and therefore $a_2 \to a_1 \mu^+ \mu^-$).

\begin{figure}[ht!]
 \begin{center}
  \includegraphics[width=\columnwidth]{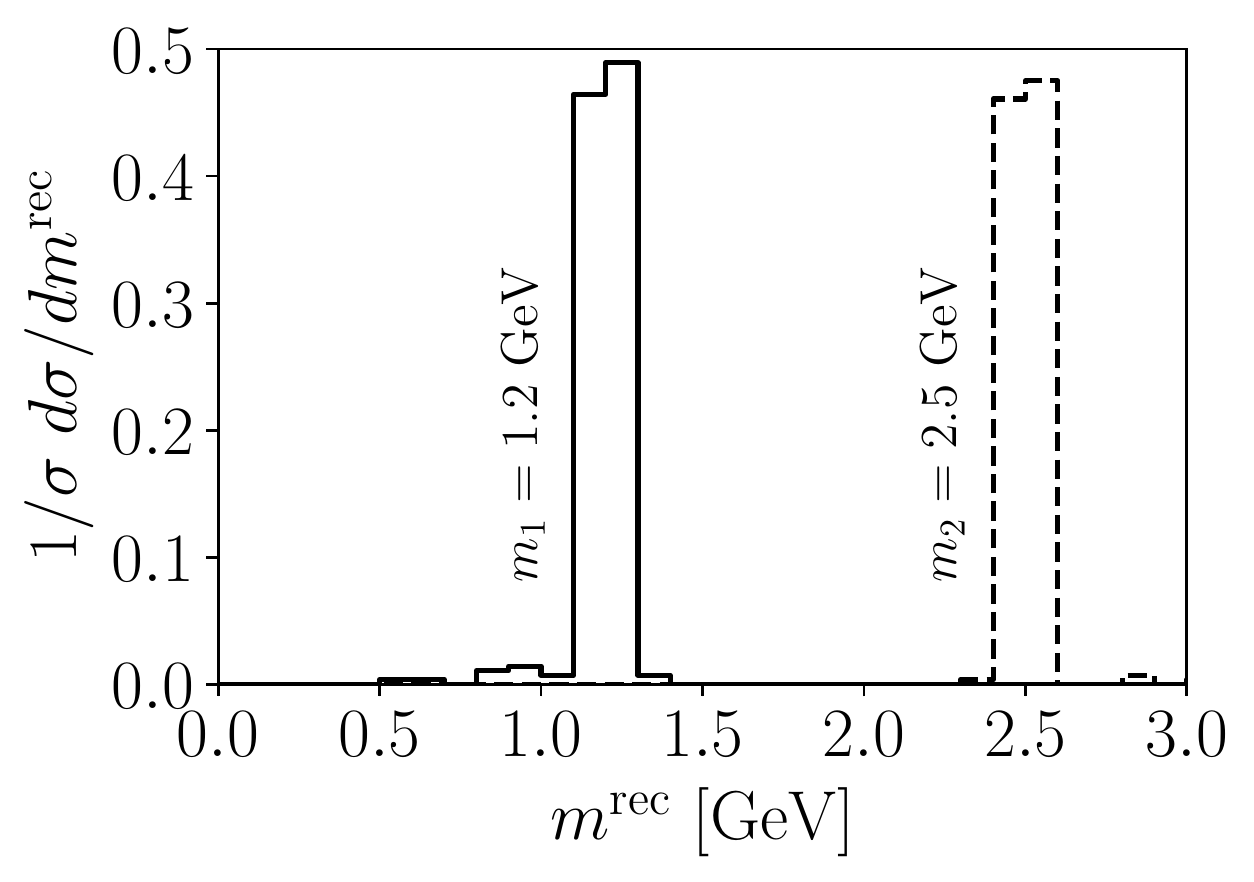}
 \end{center}
\caption{\it Normalized distribution of the reconstructed $m_1$ (solid) and $m_2$ (dashed) for $m_2 > 2m_1$.}
\label{fig:m11_R1}
\end{figure}

For the first case, we minimize the difference $|m_{11}^{\rm rec} - m_{12}^{\rm 
rec}| + |m_{12}^{\rm rec} - m_{13}^{\rm rec}|$, where $m_i^{\rm rec}$ is the 
invariant mass of each combination of opposite-sign muons. The two $a_1$s that 
reconstruct the heavier scalar are those with the minimum $\Delta R$ among 
themselves; see Fig.~\ref{fig:m11_R1} for an example. 

\begin{figure}
 \begin{center}
  \includegraphics[width=\columnwidth]{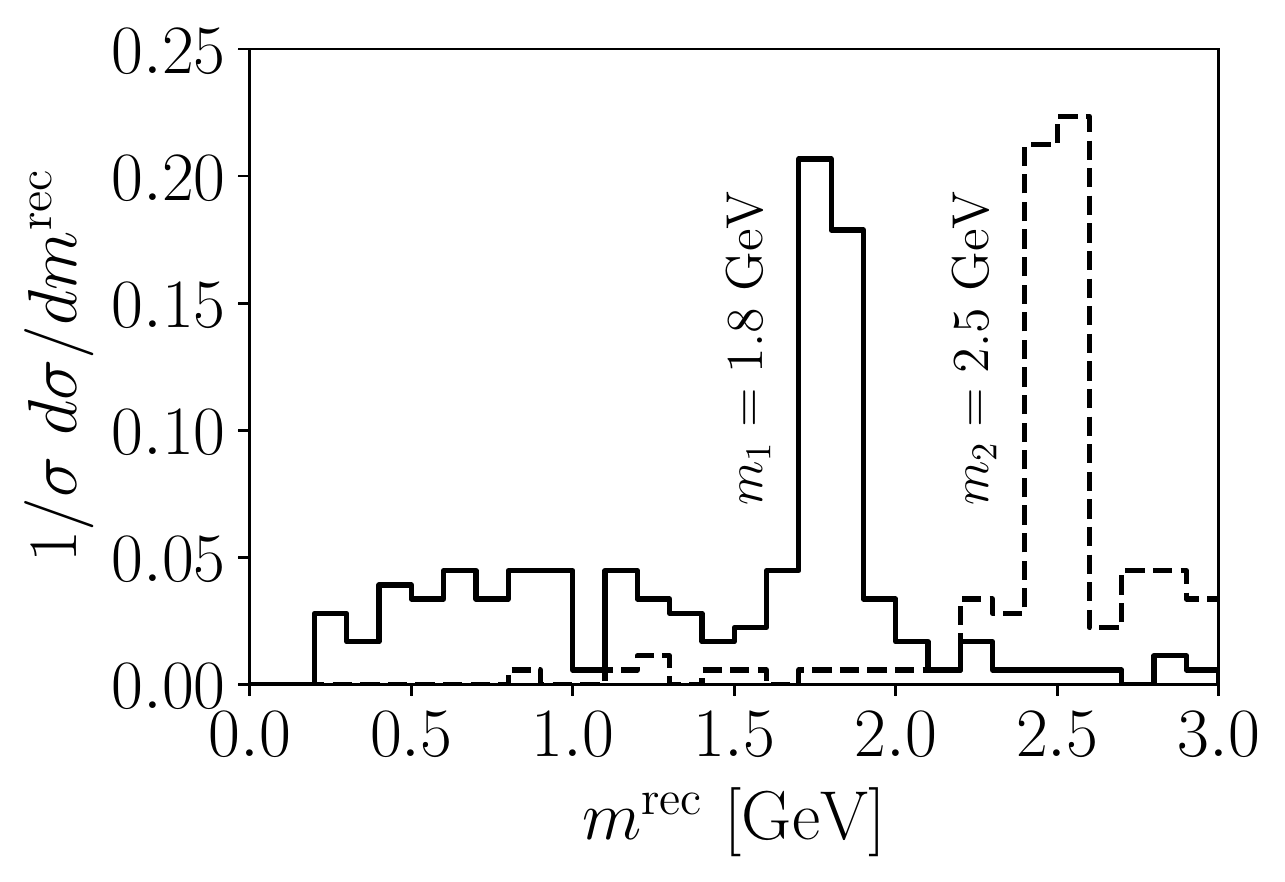}
 \end{center}
\caption{\it Normalized distribution of the reconstructed $m_1$ (solid) and $m_2$ (dashed) for $m_2 < 2m_1$.}
\label{fig:m11_R2}
\end{figure}

Concerning the second case, the muon pairs reconstructing the two $a_1$s are 
selected as those minimizing the difference $|m_{11}^{\rm rec} - m_{12}^{\rm 
rec}|$ among the three pairs of muons. Then, 
$a_2$ is reconstructed from the two muons not assigned to any $a_1$ and the $a_1$ that minimizes $\Delta R  (p_{1}, p_{\mu \mu})$ (with $p_1$ its 
four-momentum and $ p_{\mu 
\mu}$ the four-momentum of the aforementioned pair of muons); see Fig.~\ref{fig:m11_R2}.

\section{High mass regime at the LHC}\label{sec:lhc}
In the high mass regime, $a_{1,2}$ can no longer be produced in the decay of 
$B$ mesons. However, if $V$ is light enough ($m_V \lesssim $ few TeV), it can 
be produced on shell at colliders, giving rise to $a_{1,2}$ pair production 
upon decay. The tree level signal 
cross section 
for $g_{qq} = 0.5$ and $g_{12} = 1$ ranges between $\sim 0.04$ pb and 
$\sim 10^{-5}$ pb for $m_V$ between $1$ and $5$ TeV.

There are multilepton searches at the LHC which are very sensitive to this 
scenario. Most of them rely on substantial missing energy, being therefore not 
relevant for our model. In 
this work, we consider the signal region dubbed SR0A in the analysis of 
Ref.~\cite{Aaboud:2018zeb}. The main selection cuts of that study 
are \textit{(i)} at least 
four isolated leptons; \textit{(ii)} no hadronic taus; \textit{(iii)} no pair of 
opposite-sign leptons with invariant mass in the range $[81.2, 101.2]$ GeV; 
\textit{(iv)} $m_\text{eff} > 600$ GeV, where $m_\text{eff}$ stands for the 
scalar sum of the $p_T$ of all leptons, jets with $p_T^j > 40$ GeV and missing energy.

Only hadronic tau candidates with $p_T^\tau > 20$ GeV are considered in 
\textit{(ii)}; jets are reconstructed 
using the anti-$k_t$ algorithm with $R=0.4$. The experimental analysis reports 
the observation of 13 events, 
while $10.2\pm 2.1$ are predicted in the background-only hypothesis. Using 
these numbers including the systematic uncertainty on the SM prediction, we 
obtain that the maximum number of allowed signal events  is 
$12$. Scaling the expected number of background events with the larger 
luminosity, and assuming the same uncertainty, the expected maximum number of 
signal events at the HL-LHC  is $300$.

We recast this analysis using homemade routines based on 
\texttt{ROOT v5}~\cite{Brun:1997pa}, \texttt{HepMC v2}~\cite{Dobbs:2001ck} and \texttt{FasJet 
v3}~\cite{Cacciari:2011ma}. We define hadronic taus as jets with angular separation smaller 
than $0.2$ from a true hadronic decayed tau lepton. We establish a flat 
tau-tagging efficiency of $0.5$. We consider light leptons to be isolated if 
the hadronic activity around $\Delta R=0.2$ of the corresponding lepton is 
smaller than 10\% of its transverse momentum. On top of the cuts 
above, we require that 
the angular separation between any pair of muons is larger than $0.05$, to 
simulate their correct reconstruction at detectors. 

We 
generate signal events for $pp\to V\to a_1 a_2$ with the corresponding scalar 
decays with \texttt{MadGraph v5}~\cite{Alwall:2014hca} with no parton level 
cuts. For the 
PDFs we use the \texttt{NNPDF23LO} set~\cite{Ball:2012cx}. Signal events are 
subsequently passed through \texttt{Pythia v8}~\cite{Sjostrand:2014zea} to 
account for initial 
and final state radiation, fragmentation and hadronization effects. 

\begin{figure}[t]
\includegraphics[width=\columnwidth]{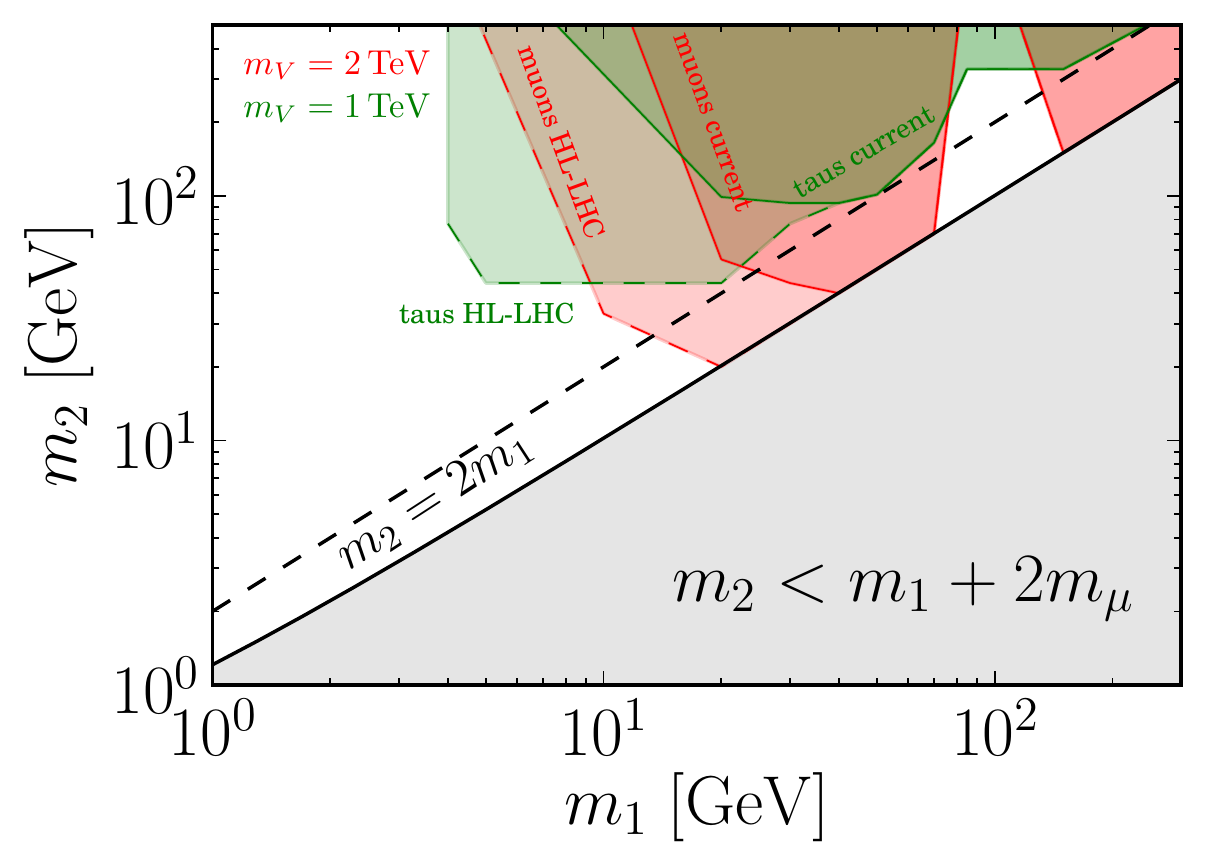}
 \caption{Region in the plane $(m_1, m_2)$ that is excluded by 
multilepton searches (solid red) 
and lepton-tau searches (solid green)~\cite{Aaboud:2018zeb}. The dashed lines 
represent the 
corresponding prospects at the HL-LHC. We have fixed $g_{qq} = 0.5$, $g_{12} = 
1$.}\label{fig:LHCmass}
\end{figure}

\begin{figure}[t]
\includegraphics[width=\columnwidth]{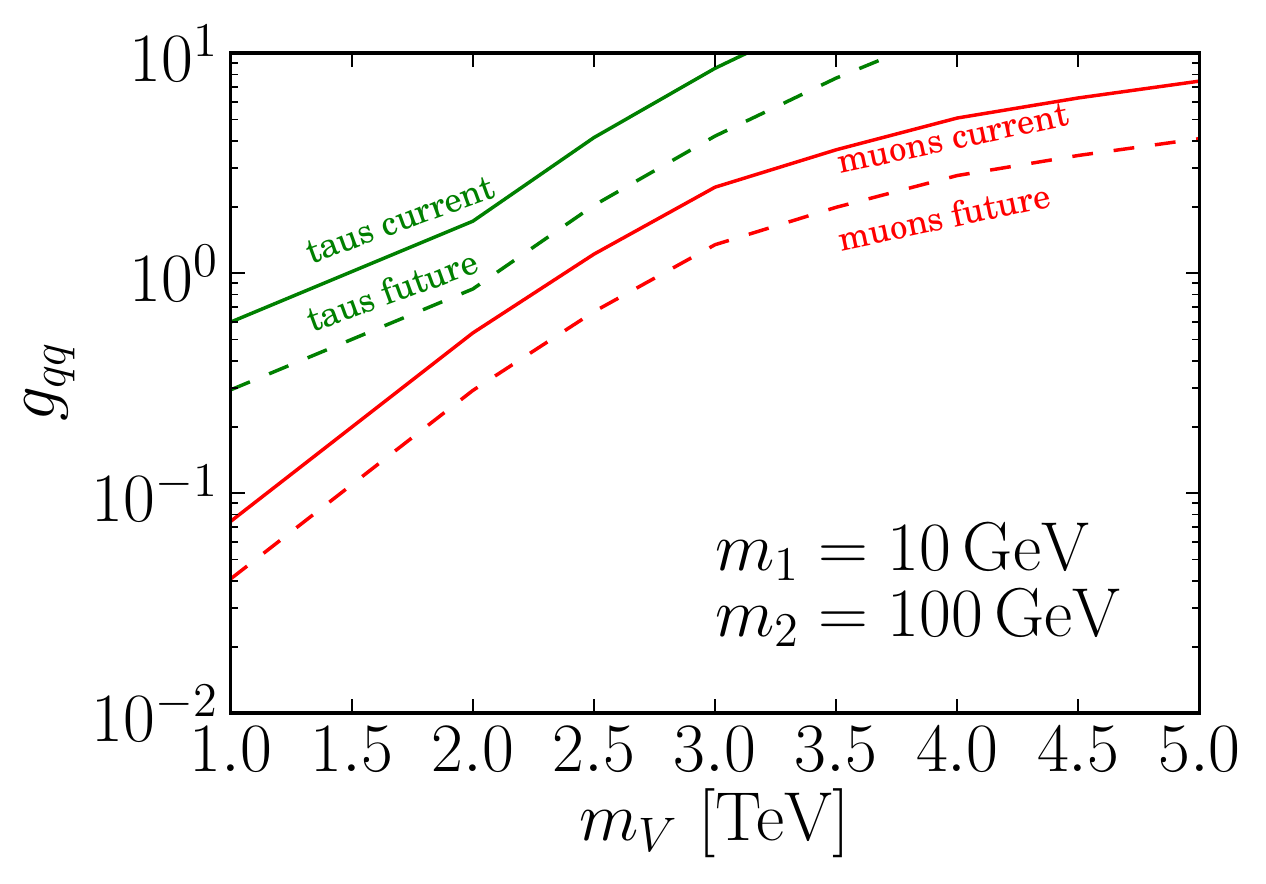}
 \caption{Minimum value of $g_{qq}$ that it is excluded  by 
multilepton searches (solid red) and 
lepton-tau searches (solid green)~\cite{Aaboud:2018zeb}. The dashed lines 
represent the 
corresponding prospects at the HL-LHC. We have assumed $\mathcal{B}(V\to a_1 
a_2)\sim 0.25$.}\label{fig:LHCV}
\end{figure}

If the light scalars couple mostly to the tau lepton (second scenario 
introduced in Section~\ref{sec:framework}), the aforementioned signal region 
has no sensitivity. We can rely instead on the signal region SR2 defined in the 
same experimental paper of Ref.~\cite{Aaboud:2018zeb}, which requires \textit{(i)} exactly 
two light leptons with invariant mass not in the range $[81.2, 101.2]$ GeV; 
\textit{(ii)} at least two hadronic taus with $p_T^\tau > 30$ GeV; 
$m_\text{eff} > 650$ GeV. The experimental collaboration reports the 
observation of 2 events; the SM prediction being $2.3\pm 0.8$. Using again the 
CL$_s$ method, we obtain 6 (121) events as the current (future) maximum 
allowed signal.

We scan over 20 values of $m_1$ and $m_2$ in logarithmic scale in the range 
$[1, 500]$ GeV, with special attention to low masses as well as masses close to 
the $Z$ pole. 

In Fig.~\ref{fig:LHCmass} we depict the region in the $(m_1, m_2)$ plane 
for $g_{qq} = 0.5$ and $g_{12} = 1$ that is already excluded in the muonphilic case and also in the case with couplings to taus. The exclusion prospects for the HL-LHC, defined by 3 ab$^{-1}$, are
also shown. The tau analysis is much less 
constraining (mainly due to the small branching ratio to leptons), and thus we only show 
results for $m_V = 1$ TeV.

The low sensitivity in the small $m_1$ region is due to muons being very 
collimated. (Decays into taus are furthermore forbidden for $m_1 \lesssim 4$ 
GeV.) If it were possible to resolve muons with angular separations as 
small as $0.001$, then almost the whole small mass range could be tested in the 
muonphilic case.

Likewise, the non excluded region around $m_1\sim 100$ GeV results from the $Z$ 
veto of the analysis. This region could be covered if the veto on the $Z$ pole 
is removed and, instead of $m_{\text{eff}}$, the invariant mass of all final state observable objets (which in our signal, 
and contrary to the SUSY targets of the analysis, presents a narrow peak) is 
used. Such improvement would also extend the reach 
to smaller masses. It is therefore desirable that future updates of the 
experimental work consider different versions of the cut on 
$m_{\text{eff}}$.

In the same vein, in Fig.~\ref{fig:LHCV} we plot the minimum value of $g_{qq}$ 
that can be tested for different values of $m_V$ and for fixed values of 
$m_{1,2}$. We have also fixed $g_{12}$ to the value for which 
$\mathcal{B}(V\to a_1 a_2)\sim 0.25$.

\section{Conclusions}\label{sec:conclusions}

We have studied the phenomenology of light leptophilic scalars $a_{1,2}$ that
couple to a heavy flavour violating (mostly $b-s$ like) spin-1 resonance $V$. 
We have shown that, under very mild conditions, $a_2$ decays mostly into $a_1$, which subsequently
decays into pairs of leptons. Thus, for scalar masses $\lesssim$ few GeV,
this scenario produces new $B$ meson decays into six muons, namely $B_s^0\to 3\mu^+ 3\mu^-$
and $B^+\to K^+ 3\mu^+ 3\mu^-$. Interestingly, the later dominates over the second
when $m_1\sim m_2$. None of them has been explored experimentally; we have
therefore proposed dedicated analyses to explore these signals at the LHCb.
We have found that branching ratios as small as $6.0\times 10^{-9}$ ($5.9\times 10^{-9}$) 
for the first (second) process
can be already tested with the current luminosity. Branching ratios hundred times smaller 
could
be probed at the Upgrade-II of the LHCb.

For larger scalar masses, $a_{1/2}$ arise rather in the decay of $V$,
which can be produced on shell at $pp$ collisions at the LHC. Current multilepton
searches in final states with muons (taus) constrain most of the parameter 
space for $m_1\gtrsim 10$ GeV provided that $\sigma(pp\to V\to a_1 
a_2)\gtrsim 0.001\, (0.01)$ pb. Smaller masses give rise to very collimated 
leptons (or jets) that are difficult to disentangle at detectors. However, at 
the HL-LHC,
the reach can be extended to $m_1\lesssim 5$. And even further if the 
current analyses
cut on the invariant mass of all visible objets.

Finally, let us comment how these results would get modified if different
flavour assumptions are made. To start with, if $a_{1,2}$ are not leptophilic
but rather they couple to all SM fermions with Yukawa-like couplings, the
branching ratio of $a_1$ into leptons would get reduced by one-to-two orders of magnitude. In 
turn,
LHCb would be only sensitive to exotic branching ratios thousand times larger. (Note that 
such branching ratios are not excluded by any current measurement, though.) However, LHC 
searches in multilepton final states
would lose almost all sensitivity in this case.

On the other hand, $V$ might also induce $b-d$ transitions. In that case, we
expect new rare decays such as $B^0\to 3\mu^+ 3\mu^-$. The production cross
section for $B^0$ is $\sim 3.7$ larger than for $B_s^0$~\cite{Aaij:2011jp}, 
from where we estimate that
$\mathcal{B}(B^0\to 3\mu^+ 3\mu^-) \gtrsim$ $1.6\times 10^{-9}$ ($\sim 10^{-11}$) can 
be probed 
currently (in the Upgrade-II of the LHCb).

On the theory side, this channel vanishes also at tree level when
$m_1\sim m_2$. In this regime, we propose searching for $B_s^0\to K^{*0} 3\mu^+ 3\mu^-$,
with $K^{*0}\to K^+ \pi^-$; whose branching ratio is around $2/3$~\cite{Tanabashi:2018oca}. 
Upon 
performing
an equivalent analysis to that described in Sec.~\ref{sec:lhcb}, we obtain 
efficiencies of about $2$ times smaller, in comparison to the $B_s^0 \rightarrow 3\mu^+ 3\mu^-$ channel. Consequently, we estimate the LHCb reach to be
$\mathcal{B}(B_s^0\to K^{*0}3\mu^+ 3\mu^-)\gtrsim$ $1.8\times 10^{-8}$  currently, and again about hundred times stronger in 
the Upgrade-II.

At high scalar masses, the prospects are only slightly better than for $b-s$
transitions, because the production cross section for $V$ at the LHC grows only
by a very small factor. Both low and high energy searches are also more 
constraining
than bounds on $\Delta M_d$~\cite{Foldenauer:2016rpi} on a wide region of 
the parameter space.

Overall, our study motivates new searches for $B_s^0\to (K^{0*}) 3\mu^+ 3\mu^-$ 
and
$B^+\to K^+ 3\mu^+ 3\mu^-$ at the LHCb as well as small modifications of current 
multilepton and multitau analyses at CMS and ATLAS.

\section*{Acknowledgments}
\noindent
We are grateful to Ulrik Egede for previous collaboration that opened this new line of research.
MC is supported by the Royal Society under the Newton International Fellowship 
programme. MR is supported by Funda\c{c}\~ao para a Ci\^encia e Tecnologia (FCT)
under the grant PD/BD/142773/2018 and also acknowledges financing from LIP 
(FCT, COMPETE2020-Portugal2020, FEDER, POCI-01-0145-FEDER-007334). MR would 
like to thank the IPPP Durham for hospitality where the main part of this word 
was carried out. MS acknowledges the hospitality of the University of Tuebingen and support of the Humboldt Society during the completion of parts of this work.

\noindent

\appendix

\section{Concrete composite Higgs model}
\label{sec:app}
\begin{figure}[t]
 \begin{center}
  \includegraphics[width=0.95\columnwidth]{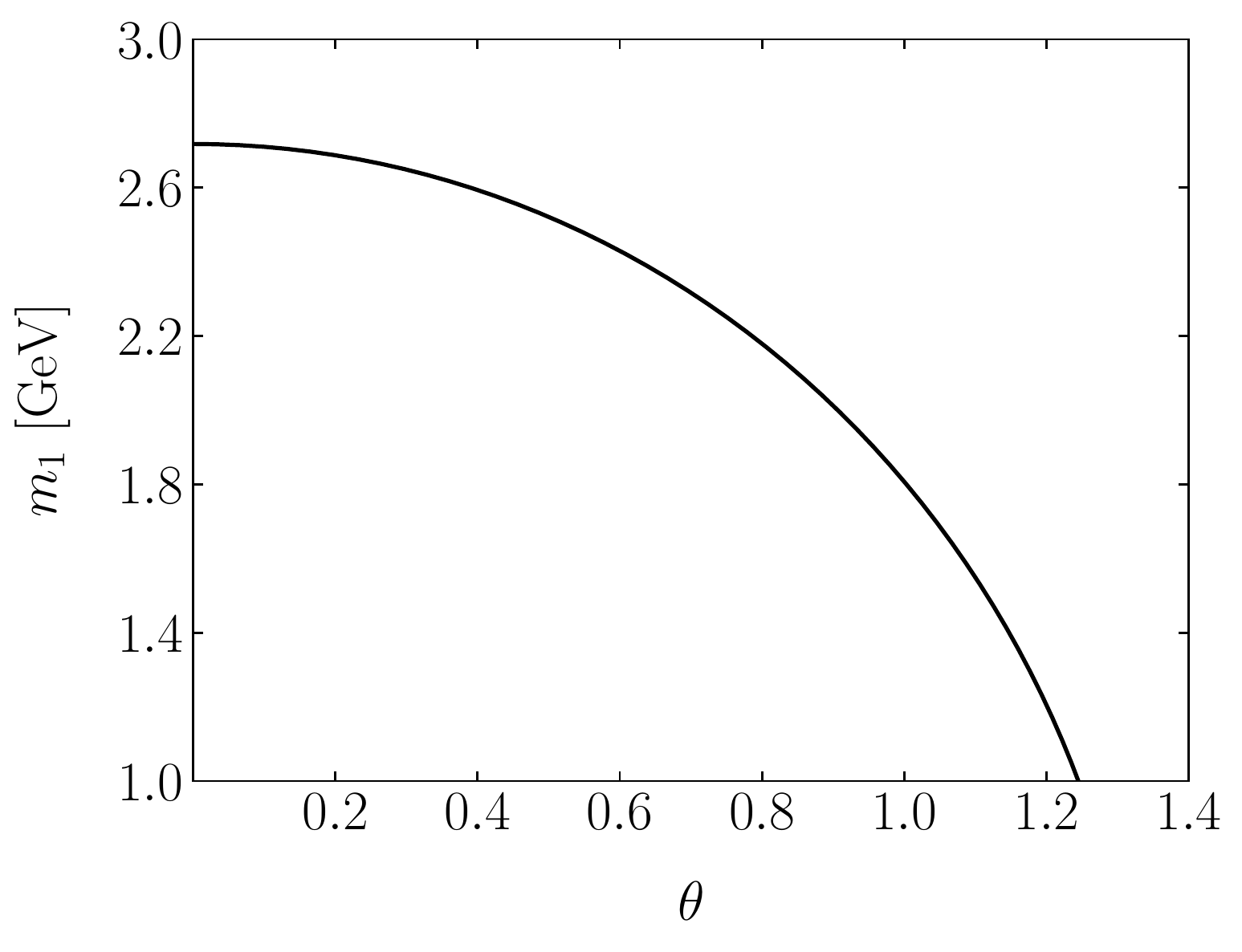}
 \end{center}
\caption{\it Mass of $a_{1}$ as a function of $\theta$ (for $f = 1$ TeV), obtained from the 
embedding of the left-handed leptons in the symmetric representation of $SO(7)$.}
\label{fig:m1,m2}
\end{figure}
Non-minimal CHMs is the context where heavy vector bosons and new light 
scalars, separated by a large mass gap, arise more naturally. The reason is that 
the latter are pseudo Nambu Golstone bosons (pNGBs) from the
the spontaneous breaking of $\mathcal{G}/\mathcal{H}$, at a 
scale $f \sim$ TeV.

The smallest coset for which the scalar sector consists of the Higgs degrees of 
freedom as well as two SM singlets is $SO(7)/SO(6)$~\cite{Chala:2016ykx,Balkin:2017aep,DaRold:2019ccj}. The corresponding 
$15$ unbroken 
and $6$ broken generators, $T$ and $X$ respectively, can be 
written as:
\begin{eqnarray} 
& T_{ij}^{mn} & = -\frac{i}{\sqrt{2}} \left(\delta_i^m \delta_j^n - 
\delta_i^n \delta_j^m \right)~, \quad m <n \in [1, 6]~; \nonumber \\
&& \nonumber\\
& X_{ij}^{mn} & = -\frac{i}{\sqrt{2}} \left(\delta_i^m \delta_j^7 
- 
\delta_i^7 \delta_j^m \right)~, \quad m \in [1, 6]~.
\end{eqnarray}

Without loss of generality, the pNGB matrix can be written as
\begin{widetext}
\begin{equation}
 U = 	\left[\begin{array}{cccccc}
 \mathbf{1}_{3\times 3} & & & \\[0.1cm]
 					 & 1 - h^2/(f^2 + f^2\Sigma) & 
-h a_1/(f^2 + f^2\Sigma) & -h a_2/(f^2 + f^2\Sigma) & 
h/f \\[0.1cm]
 					 & -h a_1/(f^2 + f^2\Sigma) & 1-  
a_1^2/(f^2 + f^2\Sigma) &- a_1 a_2/(f^2 + f^2\Sigma) & 
a_1/f \\[0.1cm]
 					 &- h a_2/(f^2 + f^2\Sigma) & - 
a_1 a_2/(f^2 + f^2\Sigma) & 1 -  a_2^2/(f^2 + f^2\Sigma) & 
a_2/f \\[0.1cm]
 					 & - h/f & - a_1/f & 
-a_2/f & \Sigma \\
	\end{array}\right]~,
\end{equation}
\end{widetext}
with $\Sigma^2 = 1-(h^2 + a_1^2 + a_2^2)/f^2$.

Following the partial compositeness paradigm~\cite{Kaplan:1991dc}, the couplings of 
$a_{1,2}$ to the SM fermions, as well as the scalar potential, depend on the 
quantum numbers of the composite operators that the SM fermions mix with 
breaking the global symmetry. Or equivalently, they depend on how the SM fermions are embedded in 
representations of $SO(7)$. 

We assume that $q_L+u_R\sim\mathbf{7}+\mathbf{21}$. Likewise, we assume that 
$l_L+e_R\sim \mathbf{27}+1$. Explicitly, $L_L\equiv\nu_L\Lambda^e + e_L 
\Lambda^\nu$: 
\begin{equation}
 L_L = 	
\frac{1}{2}\left(\begin{array}{cccc}
              0_{4\times 4} & \theta\mathbf{v_1}^T & \gamma\mathbf{v_2}^T & 
\mathbf{v_2}^T\\
              \theta\mathbf{v_1} & 0 & 0 &0\\
              \gamma\mathbf{v_2} & 0 & 0 &0\\
              \mathbf{v_2}^T & 0 & 0 & 0
             \end{array}\right)~,
\end{equation}
where the vectors read $\mathbf{v_1} = (e_L, -ie_L, \nu_L, i\nu_L)$ and 
$\mathbf{v_2} = (ie_L, e_L, 
i\nu_L, -\nu_L)$ and $\theta$ and $\gamma$ are real parameters.
(Note that the different embeddings for quarks and leptons is primarily 
justified by the fact that the lepton and quark masses and mixings are 
completely different.)

The scalar potential can be written as $V(h,a_{1,2}) = V_q(h,a_{1,2}) + V_l(h, 
a_{1,2})$, where the first and second contributions of the RHS come from loops 
of quarks and leptons, respectively.
It can be also shown that the quark sector respects a symmetry $a_{1,2}\to 
-a_{1/2}$, as well as the shift symmetry of the singlets. 
Consequently, 
$V_q(h,a_{1,2}) = V_q(h)$. It is completely fixed 
by the measurements of the Higgs mass and its VEV.

The only model dependence come from $V_l(h, a_{1,2})$, which to leading order 
in the expansion in the global symmetry breaking parameters reads:
\begin{align*}
 V_l &\sim c_1  f^4 \bigg[\left(\Lambda_D^{\mathbf{1 *}} \right)^\alpha 
\left(\Lambda_D^{\mathbf{1}} \right)_\alpha \bigg] + 
c_2  f^4 \bigg[\left(\Lambda_D^{\mathbf{6 *}} \right)^\alpha_i 
\left(\Lambda_D^{\mathbf{6}} \right)^i_\alpha	\bigg]~,
\end{align*}
where the dressed spurion reads $\Lambda_D^\alpha \equiv U^T 
\Lambda^\alpha U$ with $\alpha = e,\nu$.  (The indices $\mathbf{1}$ and $\mathbf{6}$ indicate the projection into the singlet and the sextuplet in the decomposition $\mathbf{27}=1+\mathbf{6}+\mathbf{20}$ from $SO(7)$ to $SO(6)$.)
The constants $c_1$ and 
$c_2$ are free parameters encoding the (unknown) details on the strongly 
coupled UV. 
Writing explicitly the one-loop induced potential, we find:
\begin{align}\nonumber
 V_l & =   4  f^3 c_2 \gamma \Sigma a_2  + 2 f \left(c_1 - 2 c_2\right) \gamma \Sigma a_2 h^2\\\nonumber
&+   \frac{1}{2} c_2f^2 \bigg[\left(\gamma^2 + \theta^2 - 7 + 2 \frac{c_1}{c_2} \right) h^2\\
& + 4 \left(\theta^2 - 1\right)a_1^2 + 4 \left(\gamma^2 -1\right) a_2^2 \bigg] \nonumber \\
&+ \left(c_1 - 2 c_2\right)\bigg[\left(\theta^2 - 1\right)a_1^2 + \left(\gamma^2 -1\right) a_2^2 - h^2 \bigg] h^2~.
\end{align}
We further expand this expression in powers of $1/f$, and keep 
only terms up to dimension four: 
\begin{align}\nonumber
 V_{l} &  \sim  4  f^3 c_2 \gamma a_2 + 2 f^2  c_2\bigg[\left(\gamma^2 - 1\right) a_2^2 + \left(\theta^2 -1\right) a_1^2 \bigg]\\
& + 2 f \gamma\bigg[\left(c_1 - 3c_2\right) a_ 2 h^2 - c_2  \left(a_1^2 + a_2^2\right)a_2 \bigg] \nonumber \\
& + (c_1 - 2 c_2) \bigg[ \left(\theta^2 - 1\right) a_1^2 + \left(\gamma^2 - 1 \right) a_2^2  \bigg] h^2 + ...
\end{align}
where the three dots encode terms involving the Higgs boson solely. 

The requirements $c_1\sim 3 c_2$ and $\gamma \sim 1$ make the interactions between $a_{2}$ and the Higgs (in particular mixings) very small. 
In order to avoid bounds from Higgs searches, we restrict to this case hereafter. The tadpole can then be removed by the field redefinition $a_2 \rightarrow a_2 + \sqrt{2/3} f$.

Let us also fix $f=1$ TeV, as well as $c_2\sim  g_*^2 y_l^2 / (4\pi)^2\sim 10^{-6}$. The latter is the value expected from SILH power counting~\cite{Giudice:2007fh} for $g_*\sim 3$, with $g_*$ the typical strong coupling between composite resonances. 
This choice fixes both $m_2$ and $m_{12}$ to $\sim 3.1$ GeV and $\sim 0.002$ GeV, respectively; while $m_1$ depends solely on $\theta$. We compute numerically this dependence and it is depicted  in Fig.~\ref{fig:m1,m2}.

On another front, the Yukawa Lagrangian to dimension five reads:
\begin{align}
 L_Y &= y f \overline{l_L}^\alpha 
\left(\Lambda_D^\alpha\right)^\dagger_{77} e_R + \text{h.c.} \nonumber 
\\
 &= y_\ell \overline{l_L} H e_R 
\left[1+\frac{1}{f}(a_2 - i\theta a_1)+\cdots\right]~.
\end{align}
The vector resonance associated to the generator $T^{56}$ is the only one that couples to $a_{1,2}$. We identify it with $V$.
The interaction between $V$ and the pNGBs is entirely determined by the CCWZ 
formalism~\cite{Coleman:1969sm,Callan:1969sn,Panico:2015jxa}, reading:
\begin{equation}
\label{eq: Lchm}
L = \frac{1}{2g_*^2}m_V^2 \left(g_* V_\mu^a - e_\mu^a \right)^2~,
\end{equation}
where $e_\mu$ is the trace of the 
Maurer-Cartan form $\omega$ along the unbroken generators:
\begin{equation}
 \omega_\mu = -iU^\dagger\partial_\mu U = d_\mu^a X_a + e_\mu^a T_a~.
\end{equation}

We expect $m_V \sim g_* f$. Therefore, the 
interaction between the vector resonance and the light scalars reads: 
\begin{equation}
L _{int} = \sqrt{2}\,g_* \frac{ V_\mu a_2 
\overleftrightarrow{\partial^\mu} a_1}{1+ \Sigma}  \sim \frac{g_*}{\sqrt{2}} V_\mu a_2 
\overleftrightarrow{\partial^\mu} a_1~.
\end{equation}
Finally, the vector resonance can not couple directly to the left-handed quarks. The 
coupling $g_{qq}$ is therefore suppressed by $v^2/f^2\lesssim 0.1$.

Altogether, this model matches into the parameterization in Eq.~\ref{eq:lag}. 
For example, let us take $\theta = 1.2$. We obtain: $m_1 \sim 1.3$ GeV, $m_2\sim 3.1$ GeV, $m_{12}\sim 0.002$ GeV, $g_{qq}\sim 0.1$, $g_{12}\sim 2$, $g_2 \sim 0.17$, $g_1\sim 0.22$.

These numbers are also obtained if the leptons are embedded in $\mathbf{7}+\mathbf{7}$.

\bibliographystyle{style}
\bibliography{notes}{}

\end{document}